\documentclass[superscriptaddress,twocolumn,showpacs,preprintnumbers,amsmath,amssymb,pra]{revtex4}

\usepackage{graphicx,epsfig,psfrag}
\usepackage{dcolumn}
\usepackage{bm}

\newcommand{\bfr}{{\bf r}}

\begin{document}

\title{Superfluidity of an interacting trapped quasi-2D Bose gas}
\author{T.~P. Simula}
\affiliation{Lundbeck Foundation Theoretical Center for Quantum System Research,\\
Department of Physics and Astronomy, University of Aarhus, DK-8000 Aarhus C, Denmark}
\affiliation{Laboratory of Physics, Helsinki University of Technology, P.O.Box 1100, FI-02015 TKK, Finland }
\author{M.~J. Davis}
\affiliation{ARC Centre of Excellence for Quantum-Atom Optics, School of Physical Sciences, \\
Department of Physics, University of Queensland, Brisbane, QLD 4072, Australia }
\author{P.~B. Blakie}
\affiliation{Jack Dodd Centre for Photonics and Ultra-Cold Atoms, Department of Physics, University of Otago, New Zealand}

\date{\today}

\begin{abstract}
We investigate the harmonically trapped interacting Bose gas in a quasi-2D geometry using the classical field method. The system exhibits quasi-long-range order and non-classical rotational inertia at temperatures below the Berezinskii-Kosterlitz-Thouless cross-over to the superfluid state. In particular, we compute the scissors-mode oscillation frequencies and find that the irrotational mode changes its frequency as the temperature is sweeped across the cross-over thus providing microscopic evidence for the emergence of superfluidity.
\end{abstract}

\pacs{03.75.Lm, 67.40.Vs}
\maketitle

\section{Introduction}

Superfluids are peculiar states of matter in which, at the cost of losing part of their individuality, particles gain the ability of cooperative lossless motion \cite{LeggettBook}. The occurrence of superfluidity can be attributed to the formation of certain non-local correlations within the system. For two-dimensional (2D) systems, Mermin and Wagner showed that Heisenberg models can be neither ferromagnetic or anti-ferromagnetic at finite temperatures \cite{Mermin1966a}. Hohenberg further ruled out the existence of long-range ordering in 2D Bose and Fermi systems \cite{Hohenberg1967a}. These rigorous results imply that Bose-Einstein condensation (BEC) does not exist at any finite temperature in uniform, interacting 2D systems in the thermodynamic limit, since spontaneous long-range ordering is prevented by long-wavelength fluctuations. Hence one might expect that conventional superfluidity would not occur in 2D systems. 

Nevertheless, a different path to superfluid behavior is possible in 2D systems. At low temperatures quasi-long-range correlations may form with an associated power-law decay that eventually reaches zero instead of extending throughout the system. It was theoretically shown by Berezinskii \cite{Berezinskii1971a} and by Kosterlitz and Thouless \cite{Kosterlitz1973a} (BKT) that in 2D a transition to a superfluid state may occur at finite temperatures. Qualitatively the physics of such low-temperature 2D system is conveniently described using the notion of vortex-antivortex pairs (VAPs). At temperatures below the BKT transition, long-wavelength fluctuations destroy true long-range order and yield spontaneous creation and annihilation of bound VAPs at the boundaries of the local domains of the resulting ``quasicondensate'' \cite{Popov1983a}. These locally coherent blocks contribute to the power-law decay of the two-point correlation function resulting in a superfluid response of the system. On increasing the temperature, fluctuations increase and VAPs unbind at the BKT transition. The breaking of VAPs results in the proliferation of free vortices and an exponentially decaying correlation length, and hence the system loses its superfluid properties. This was further quantified by Nelson and Kosterlitz who predicted a universal jump in the superfluid density at the critical point, which may be used to empirically detect the BKT transition \cite{Nelson1977a}.

In a trapped ultra-cold Bose gas the situation is rather complicated due to the inhomogeneity arising from the confining potential. Bagnato and Kleppner showed that a trapped, \emph{ideal} Bose gas undergoes BEC at finite temperatures \cite{Bagnato1991a}, implying that in principle a trapped, \emph{interacting} Bose gas may exist in a coherent BEC phase and/or in a BKT-type phase. The low temperature structure of a real 2D Bose gas in a trap has therefore attracted a fair amount of theoretical discussion in the recent literature and it has been debated whether the superfluid transition in such systems is of BEC-type or BKT-type \cite{Petrov2000a,Prokofiev2001a,Prokofiev2002a,Andersen2002a,Gies2004a,Trombettoni2005a,Simula2005a,Simula2006a,Polkovnikov2006a,Holzmann2005a,Holzmann2007a, Bloch2007a}.

Experimentally, the superfluid BKT transition in a bulk was first realized in liquid helium thin films by Bishop and Reppy \cite{Bishop1978a}. Resnick \emph{et al.} reported an observation of the transition in superconducting Josephson junction arrays \cite{Resnick1981a}, followed by Safonov \emph{et al.} who measured a kink in the three-body loss rate in spin-polarized hydrogen \cite{Safonov1998a}. These experiments relied on indirect methods of observation where as in the trapped atomic gases VAPs and their dynamics can be directly imaged. The quasi-2D regime in trapped quantum degenerate gases has been approached experimentally by using tight axial confinement with the aid of optical potentials and through centrifugal expansion in rapidly rotated condensates \cite{Gorlitz2001a,Rychtarik2004a, Schweikhard2004a,Smith2005a,Stock2005a}. However, probing the details of the quasicondensation transition has only recently become experimentally accessible \cite{Hadzibabic2006a,Kruger2007a,Schweikhard2007a,Clade2007a}.

Indeed, Dalibard and co-workers observed phase defects in the interference patterns of multiple quasi-2D gases trapped in the valleys of an optical lattice \cite{Stock2005a}. These phase defects were evidently caused by unbound free vortices \cite{Stock2005a,Simula2006a}. Further observations on spatial phase correlations in the system provided evidence for the cross-over between the BKT quasicondensate and normal state \cite{Hadzibabic2006a,Kruger2007a}. While the observed correlations were shown to be consistent with the BKT quasicondensation picture, the question whether the system is superfluid or not still remained. Recently, an observation of the BKT cross-over has been achieved in a 2D lattice of Josephson-coupled BECs \cite{Schweikhard2007a} and in a single 2D dipole trap \cite{Clade2007a}.

The interplay between interactions and inhomogeneous effects arising from the trapping potential have made theoretical predictions for the low temperature phases of dilute, atomic Bose gas difficult. The main point we address in this paper is the question of superfluidity in this system at low temperatures. There is not a single observable that categorically defines superfluidity and we present results of microscopic calculations for a variety of observables, including off-diagonal long-range order (condensate), fluctuations, scissor mode dynamics and presence of vortices. 
From the combined analysis of these quantities we are able to infer a cross-over temperature, $T_{\rm sf}$, below which the system exhibits superfluidity. 
 
The paper is organized as follows: In Sec. II we discuss various measurables useful in examining the superfluid properties of our system. Our computational approach is explained in Sec. III, and the results are presented in Sec. IV followed by the concluding remarks in Sec. V.

\section{Evidence of superfluidity}

A substance which has the ability to flow without dissipation is superfluid. Although the difference between a superfluid and a classical fluid may seem intuitively clear, it is difficult, if not impossible, to find a single universal definition for superfluidity against which any material could be tested. Indeed, the complete description of superfluidity is not a single feature but a complex of phenomena \cite{Leggett1999a}. In an interacting 3D atomic gas, the formation of BEC is essentially equivalent to the emergence of a macroscopic wave function, which inherently exhibits long-range order throughout the system. Furthermore, the system attains finite superfluid fraction at the BEC transition. The situation is more subtle in 2D where the condensation process is plagued by long-wavelength phase fluctuations. In the following, we introduce measurables relevant for providing evidence of superfluidity in a quasi-two-dimensional sample of trapped ultra-cold atoms and apply these definitions to discuss the superfluidity of quasicondensates.

\subsection{Role of quantized vortices}
Quantized vortices are the hallmark of superfluids. The flow ${\bf v}_s(\bfr,t)=\hbar/m\nabla \varphi(\bfr,t)$ of a superfluid described by a macroscopic wave function with a phase $\varphi(\bfr,t)$ obeys the condition of irrotationality, $\nabla\times{\bf v}_s(\bfr,t)=0$, and the Onsager-Feynman quantization of circulation,
\begin{equation}
\oint{\bf v_s}(\bfr,t)\cdot{\rm{d}}{\bf l}=\kappa 2\pi \hbar/m,
\label{eq1}
\end{equation}
where $\kappa$ is an integer and $m$ is the mass of an atom. Therefore rotation ($\kappa\ne 0$) is only possible around the phase singularity at the core of a quantized vortex where the superfluid density vanishes. If normal fluid is present, it occupies the volume in the vortex core. The quantized vortices can therefore be seen to play a two-fold role in superfluid systems. On the one hand their coherent role is vital in enabling superfluids to rotate, while on the other hand they can be viewed as (topological) defects that are a source of incoherence causing a reduction in the superfluid fraction.

It is worth noting that the observation of a vortex alone is not a sufficient criterion  from which superfluidity may be deduced. For instance, while a persistent vortex in a zero temperature condensate is readily accepted as \emph{proof of superfluidity}, in a 2D system near BKT cross-over an observation of a transient isolated vortex would be more likely to signify unbinding of VAPs and hence \emph{loss of superfluidity}. Therefore,  special care must be taken in the treatment of a situation where both spontaneously (thermally activated) and actively (by external rotation) created vortices may exist simultaneously. A particularly vivid example of a classical vortex which has many of the characteristics of a superfluid vortex has been realized in a recent experiment in which the $2\pi$ vortex phase winding was imprinted into a non-superfluid cloud of atoms and was observed to persist for extended times due to the cancellation of spins in the diffusion process \cite{Davidson2007a}.

\subsection{Off-diagonal long-range order}
Consider a system described by the usual one-body reduced density matrix
\begin{equation}
\rho(\bfr,\bfr')=\langle \hat\Psi^\dagger(\bfr)\hat\Psi(\bfr') \rangle,
\label{eq2}
\end{equation}
where $\hat\Psi(\bfr)$ is the second quantized bosonic field operator  and the brackets denote quantum mechanical ensemble averaging. If $\rho$ has a macroscopic eigenvalue $N_0=\mathcal{O}(N)$ where $N$ is the number of particles in the system, it is said to possess Bose-Einstein condensation in the state determined by the corresponding eigenvector, also known as the macroscopic condensate wave function. 

The concept of long-range order is often encountered in the context of superfluidity. In a homogeneous Bose-Einstein condensed system, true long-range order exists in the sense that
\begin{equation}
\lim_{|\bfr -\bfr '|\to\infty }\rho(\bfr ,\bfr ')={\rm const.},
\label{eq3}
\end{equation}
where as in the normal state the off-diagonal correlations decay exponentially
with
the spatial separation
\begin{equation}
\rho({\bfr},\bfr')\sim e^{-|\bfr-\bfr'|/\xi_0},
\label{eq4}
\end{equation}
where $\xi_0$ characterizes the length scale over which the correlations decay. Although the existence of Bose-Einstein condensation does not \emph{a priori} imply the system to be superfluid, the existence of off-diagonal order in the system can be considered as a prerequisite for superfluidity. In the case of finite systems, such as trapped atomic gases, Eq.~(\ref{eq3}) generalises by understanding that the boundary of the system is mapped to infinity.

\subsection{Algebraic long-range order}

Two-dimensional systems lacking true long-range order may attain superfluidity through the Berezinskii-Kosterlitz-Thouless mechanism. In this case the correlations decay algebraically with distance
\begin{equation}
\rho(\bfr,\bfr')\sim \left(\frac{\xi_0}{|\bfr-\bfr'|}\right)^{\eta},
\label{eq5}
\end{equation}
and are characterised by the exponent, $\eta(T)=n_{\rm sf}\lambda_{\rm dB}^2$, where $n_{\rm sf}$ is the 2D superfluid density, and $\lambda_{\rm dB}$ is the thermal de Broglie wavelength. For temperatures below the BKT transition temperature ($T_{\rm{BKT}}$) the first order approximation to the critical exponent  is $\eta(T)=\frac{1}{4}T/T_{\rm BKT}$, corresponding to the universal jump at $T=T_{\rm BKT}$ in the superfluid density \cite{Nelson1977a}.

\subsection{Higher-order coherence}
Further insight into the state of the system is obtained by studying the higher-order coherence properties of the system \cite{Glauber1963}. The second-order coherence function
\begin{equation}
g^{(2)}(\bfr)= \frac{\langle\hat\Psi^\dagger(\bfr)\hat\Psi^\dagger(\bfr)\hat\Psi(\bfr)\hat\Psi(\bfr)\rangle}{\langle \hat\Psi^\dagger(\bfr)\hat\Psi(\bfr) \rangle^2}
\label{eq6}
\end{equation}
yields information about the local coherence of the Bose field, $\hat{\Psi}$, and in general for $p$th order coherence $g_{\rm c}^{(n)}(\bfr)=1$  for $n\le p$, while for the corresponding thermal state $g_{\rm t}^{(n)}(\bfr)=n!$. By decomposing the field into `coherent'  ($\Phi$) and `incoherent' ($\hat{\psi}$) parts, $\hat\Psi=\Phi+\hat\psi$, Eq.~(\ref{eq6}) becomes
\begin{equation}
g^{(2)}(\bfr)= \frac{\langle |\Phi|^4 + 4 |\Phi|^2 n_{\rm th} + 2 n_{\rm th}^2\rangle}{\langle|\Phi|^2 +n_{\rm th} \rangle^2},
\label{eq7}
\end{equation}
where $ n_{\rm th}\equiv \langle\hat\psi^\dagger\hat\psi\rangle$.
In a similar fashion
\begin{equation}
g^{(3)}(\bfr)= \frac{\langle |\Phi|^6 + 9n_{\rm th}|\Phi|^4 + 18n_{\rm th}^2|\Phi|^2 +6n_{\rm th}^3\rangle}{\langle|\Phi|^2 +n_{\rm th} \rangle^3},
\label{eq8}
\end{equation}
which essentially measures the probability of three-particle coincidences. The total three-body recombination rate  
\begin{equation}
\Gamma=-K_3\int g^{(3)}(\bfr)n_{\rm tot}(\bfr)^3\; d^3r
\label{eq9}
\end{equation}
where $K_3$ is the rate constant and $n_{\rm tot}(\bfr)=\langle\hat\Psi^\dagger(\bfr)\hat\Psi(\bfr)\rangle$ is the total density, is an experimentally measurable quantity yielding overall information of the third order coherence properties of the system. While it has been used experimentally to infer the quasicondensation transition point in helium thin films \cite{Safonov1998a}, obtaining similar information in trapped 2D gases is more problematic since the local gas density increases as the transition is crossed from normal to quasicondensed state and this partially compensates for the corresponding decrease in $g^{(3)}(\bfr)$. 

\subsection{Non-classical rotational inertia}
The moment of inertia, $I(T)$, of a superfluid about a chosen axis is reduced from its classical value, $I_{\rm cl} = mN\langle \bfr^2 \rangle$, due to the irrotational motion of superfluid matter. The temperature dependent superfluid fraction 
\begin{equation}
\frac{N_{\rm sf}}{N_{\rm tot}} = 1-\frac{I(T)}{I_{\rm cl}},
\label{eq10}
\end{equation}
where $N_{\rm tot}$ is the total particle number, is a quantity of special interest. A finite value of this macroscopic measurable may be used as an evidence of superfluidity. Microscopically, this information about superfluidity is encoded in the elementary excitation spectrum of the system. The collective scissors mode oscillation has been employed to prove that the occurrence of BEC in 3D implies superfluidity \cite{Marago2000a,Marago2001a,Zambelli2001a,Guery1999a}. Essentially, the scissors mode may be viewed as an oscillation of an ellipsoidal cloud of atoms about its semi-axis in the plane.
In the collisionless regime, a gas in a normal state has two prominent undamped scissors mode eigenfrequencies
\begin{equation}
\omega_\pm=|\omega_x\pm\omega_y|,
\label{eq11}
\end{equation}
where $\omega_x$ and $\omega_y$ are the planar trapping frequencies. In Eq.~(\ref{eq11}) $\omega_+$ corresponds to an irrotational quadrupole oscillation and $\omega_-$ is related to a classical rotational motion. If superfluid component is present, it oscillates at an additional characteristic frequency
\begin{equation}
\omega_{\rm sf}=\sqrt{\omega^2_x +\omega^2_y},
\label{eq12}
\end{equation}
whose existence thus provides a clear sign for superfluidity of the system in the collisionless regime. It is to be noted, however, that in the hydrodynamic limit $\omega_-$ becomes over-damped and both the remaining thermal mode, $\omega_+$, and the superfluid scissors mode, $\omega_{\rm sf}$, attain the same value. In such situation, the damping rate of this mode may in principle be used to reveal superfluid response, although this may prove to be difficult to achieve in practice.

The scissors mode excitations of the system are directly related to the reduced moment of inertia \cite{Guery1999a, Marago2000a, Marago2001a, Zambelli2001a} 
\begin{equation}
\frac{I(T)}{I_{\rm cl}}=(\omega_x^2-\omega_y^2)^2\frac{\int Q(\omega)/\omega^2\;{\rm d}\omega}{\int Q(\omega)\omega^2 \;{\rm d}\omega},
\label{eq13}
\end{equation}
where $Q(\omega)$ is the Fourier transform of the time-dependent quadrupole moment $Q(t)=\int xy\; n_{\rm tot}(\bfr,t)\;{\rm d}\bfr $. 
Substitution of Eq.~(\ref{eq13}) in Eq.~(\ref{eq10}), yields a formula for the superfluid fraction in terms of the scissors mode excitations. We emphasize that the presence of a superfluid scissors mode, $\omega_{\rm sf}$, implies non-classical rotational inertia, $I < I_{\rm cl}$, finite superfluid fraction, $N_{\rm sf}/N_{\rm tot}$, and hence superfluidity.

\section{Methods}
We employ the method of classical fields as detailed in Refs.~\cite{Davis2001a,Davis2001b,Davis2002a,Blakie2005a}. Essentially, this amounts to propagating the projected Gross-Pitaevskii equation
\begin{equation}
i\hbar\partial_t\Phi= -\frac{\hbar^2}{2m}\nabla^2\Phi  +V_{\rm ext}\Phi+g\mathcal{P}\{|\Phi|^2\Phi\},
\label{eq14}
\end{equation}
in time for the field, $\Phi(\bfr,t)$, restricted in the subspace determined by an energy cut-off in the harmonic oscillator basis states. The projector, $\mathcal{P}$, serves to constrain the evolution of the field within the subspace of highly occupied states. Each simulation corresponds to the evolution of a single trajectory through the phase space and therefore, in order to construct thermodynamic observables, one should ensemble average over many different but equivalent trajectories. However, when considering an equilibrium quantities, we may assume the system to be ergodic and replace such ensemble averages by time-averages over the instantaneous field configurations taken from a single trajectory. The field, $\Phi(\bfr,t)$, is normalized to the number of particles, $N_{\rm cl}$, described by the restricted basis. The total number of particles, $N_{\rm tot}=N_{\rm cl}+N_{\rm th}$, is obtained by using the semiclassical Hartree-Fock approximation for the $N_{\rm th}$ above cut-off particles as in Refs.~\cite{Davis2005a,Simula2006a} and as described below. The in-plane phase function, $\varphi(x,y,t)$, of the complex field, $\Phi(\bfr,t)$, allows for an explicit detection of the locations of vortices and antivortices. The classical field is completely described by the conserved total energy, $E_{\rm cl}$, an energy cut-off for the restricted basis, $E_{\rm cut}$, the dimensionless nonlinearity constant, $C=gN_{\rm cl}/\hbar\omega_x a_x^3$, and the harmonic trap frequencies, $\omega_x,\omega_y$ and $\omega_z$. Here the spatial length scale is $a_x=\sqrt{\hbar/2m\omega_x}$. From these simulations we can also compute the equilibrium temperature, $T$, and chemical potential, $\mu$, as described in Refs.~\cite{Davis2003a,Davis2005b}.

\subsection{Semiclassical approximation}
An inherent feature in our numerical method requires that the total particle number, $N_{\rm tot}$, must be computed \emph{a posteriori}. This is done within the self-consistent Hartree-Fock approximation by computing the particles not included in the simulated field, $\Phi$, from the semiclassical density. The form of the semiclassical integral reflects the quasi-2D nature of the trap. For the temperatures considered here, $k_BT\sim \hbar\omega_z$ and therefore several of the lowest axial oscillator states, $n_z=\{0,1,2\ldots\}$, contribute significantly to the total number of particles. However, the temperature is too low for the equipartition theorem to apply and therefore these axial levels need to be treated discretely in the semiclassical integral. 
The Hartree-Fock energy for this system is given by
\begin{eqnarray}
E^{n_z}_{\rm HF}(K,x,y) =  K+V_{\rm ext}(x,y,0)+(n_z+1/2)\hbar\omega_z\nonumber\\
\hspace*{5mm}+2\sum_mg_{n_z,m}n_{\rm th}^{n_z}(x,y) +2g_{n_z,0}\langle |\Phi(x,y)|\rangle^2,
\label{HFegy}
\end{eqnarray}
with $K$ the kinetic energy. The thermal densities are given by
\begin{equation}
n^{n_z}_{th}(x,y) =\frac{m}{2\pi\hbar^2}\int_{K_{\min}} ^{\infty}    \frac{1}{e^{(E_{\rm HF}^{n_z}(K,x,y)-\mu)/k_BT  }-1} \;{\rm d}K,
\label{2Ddens}
\end{equation}
where
\begin{equation}
K_{\min}=\max\{0,E_{\rm cut}  - V_{\rm ext}(x,y,0)-\hbar\omega_z/2 \},
\end{equation}
and $\mu$ is the chemical potential. The interaction term in Eq.~(\ref{HFegy}) contains the multilevel coupling constant 
\begin{equation}
g_{n,m}=g\int|\phi_n(z)|^2 |\phi_m(z)|^2  \;dz,
\end{equation}
which accounts for the interactions between the particles in different axial levels. We have denoted $\Phi(x,y,z)=\Phi(x,y)\phi_0(z)$ and $\phi_n(z)$ are the normalized harmonic oscillator eigenstates. Finally, the number of above cut-off particles 
\begin{equation}
N_{\rm th} = \sum_{n_z}\int n^{n_z}_{th}(x,y) \;dx dy,
\end{equation}
is obtained by integrating over the 2D densities and summing the contributions from different axial energy levels. To obtain data points for a fixed number of particles $N_{\rm tot}$ for a range of temperatures, we estimate the cutoff energy $E_{\rm cut}$, total number of particles $N_{\rm cl}$ and energy $E_{\rm cl}$ of the classical region of the system according to the prescription of Ref.~\cite{Blakie2007a}.  We then simulate this within the PGPE and calculate the temperature and total number of thermal particles using the procedure described above.  We use this knowledge to make any necessary adjustment to the initial guesses to end up with a target $N_{\rm tot}$.

\subsection{Collective excitations}
In principle we could construct an approximation to the full Green's function from our time-dependent classical field simulations allowing us to extract the collective excitation frequencies and their damping rates for the system. It would be, however, a formidable task. Furthermore, an identification of near-degenerate modes would become cumbersome. Instead, we concentrate on a specific class of excitations---the so-called scissors modes---which may be selectively excited and consequently their oscillation frequency can be individually measured from our dynamical simulations.

In order to accurately compute the scissors mode collective oscillation frequencies, an ensemble averaging over many equivalent trajectories is required. Therefore we first prepare a large set of initial field configurations by time-sampling a single equilibrium simulation.  These instantaneous field configurations are then rotated 11 degrees in the $x-y$ -plane with respect to the semi-axis of the anisotropic trapping potential. To facilitate lossless rotation of the state, we must first project the classical field into a larger eigenbasis in order to account for the increase in the energy after rotation due to the anisotropic trapping geometry. Subsequently the seed states thus prepared are propagated in time and the quadrupole moment, $Q_i(t)$, is measured. The above procedure is repeated for $S=150-200$ microstates for each temperature point. The scissors mode frequencies are then obtained from the Fourier transformation of the ensemble averaged quadrupole signal $\tilde Q(t)=\sum_{i=1}^SQ_i(t)/S$, from which also the damping rates of those modes can be estimated.

\subsection{Computational parameters}
In order to allow practical comparison with future experiments, we choose experimentally realistic system parameters which allow us to study the scissors-mode collective oscillation frequencies. The strength of particle interactions is determined by a constant, $g=4\pi\hbar^2 a/m$, and the harmonic confining potential, $V_{\rm ext}= m(\omega_x^2x^2+\omega_y^2y^2+\omega_z^2z^2)/2$, is characterized by the Cartesian frequencies $ \{\omega_x,\omega_y,\omega_z\}=2\pi\times \{10,20,4000\}$ Hz. The trap is chosen to be anisotropic in the $x-y$ -plane in order to lift the degeneracy between the quadrupole and scissors modes. However, the planar anisotropy is kept moderate, in contrast to Ref.~\cite{Hadzibabic2006a},  in order to separate the frequency of the superfluid scissors-mode from both of the two classical modes.  We consider $N_{\rm tot}=10^5$ $^{87}$Rb atoms interacting with the $s$-wave scattering length $a=5.3$ nm.

\subsection{Radial averaging}
Several results we present are generated by performing an average along elliptical trajectories in the 2D plane about the trap centre. This is done in order to utilize the full information contained in the simulated fields. Since our system does not possess cylindrical symmetry (in order to facilitate computation of scissors modes), we perform this averaging by considering elliptical shells of constant (2D) trap potential energy and average over all spatial points falling on such strips. In what follows, $r_0$, denotes the distance from origin to such ellipse along the weakly trapped $x$-axis of the trap.

\section{Results}
In this section we present our numerical results and analysis for the superfluid indicators described above. Taking the compendium of results described below, we claim that the quasicondensate studied here is superfluid. All measurables extracted from our simulations point to a cross-over temperature 
$T_{\rm sf}$ at which our system attains superfluidity.

\subsection{Fluctuations and vortices}
\begin{figure*}[t!]
\center
\includegraphics[viewport= 180 385 420 515,clip, width=17.2cm]{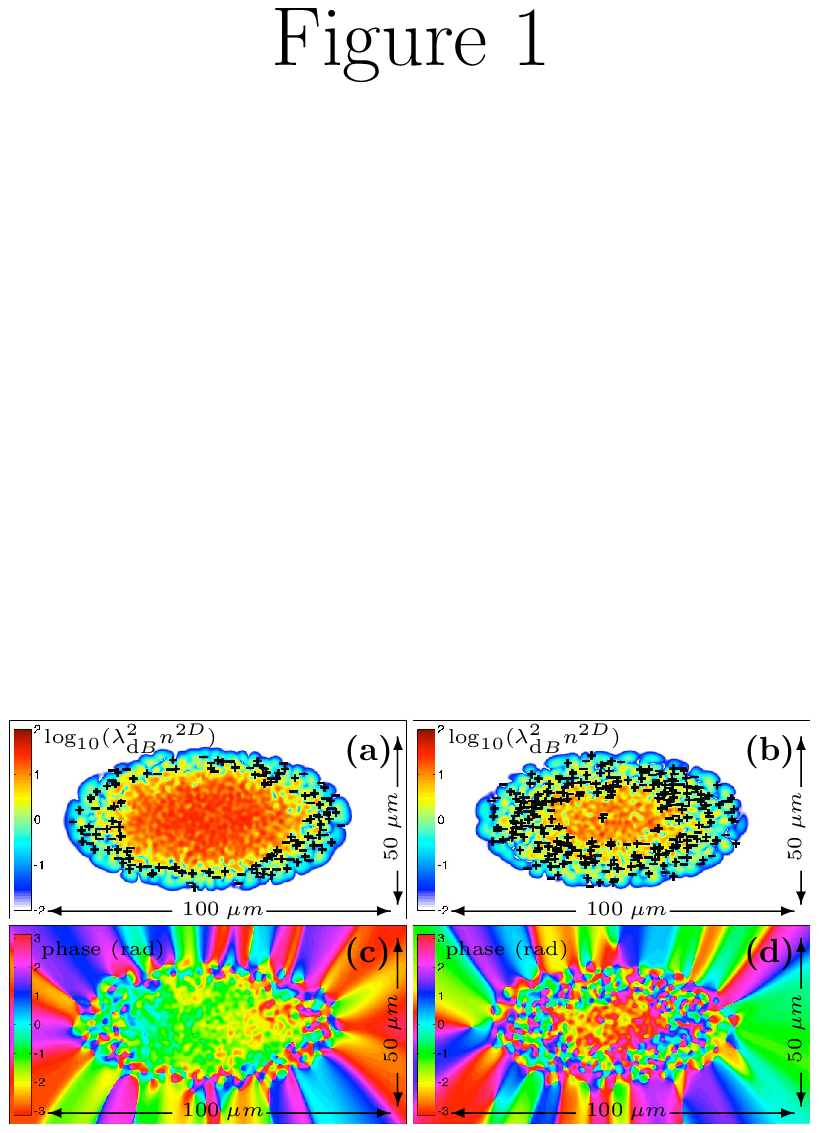}
\caption{(Color online) Density (a),(b) and phase (c),(d) of the classical field at two different temperatures: (a),(c) $T=114 \;n$K, and (b),(d) $T=151\; n$K. Vortices and antivortices are denoted by + and $-$ signs respectively.}
\label{f1}
\end{figure*}

In order to lay down the qualitative features of the system we have plotted instantaneous 2D classical field densities, $\lambda^2_{\rm dB}|\Phi(x,y)|^2$, in Fig.~\ref{f1}(a) and (b), and the corresponding phases $\varphi(x,y)$, Fig.~\ref{f1}(c) and (d). The temperatures are $T=114\;n$K for Fig.~\ref{f1}(a) and (c) and $T=151\;n$K for Fig.~\ref{f1}(b) and (d).
At low temperatures the density and phase are relatively uniform, while at high temperatures both exhibit strong fluctuation and and vortices and antivortices are nucleated. This observation is in striking contrast to the usual situation in 3D and highlights the main qualitative difference between 2D and 3D systems.

\begin{figure}
\center
\includegraphics[viewport= 180 410 430 610,clip, width=8.6cm]{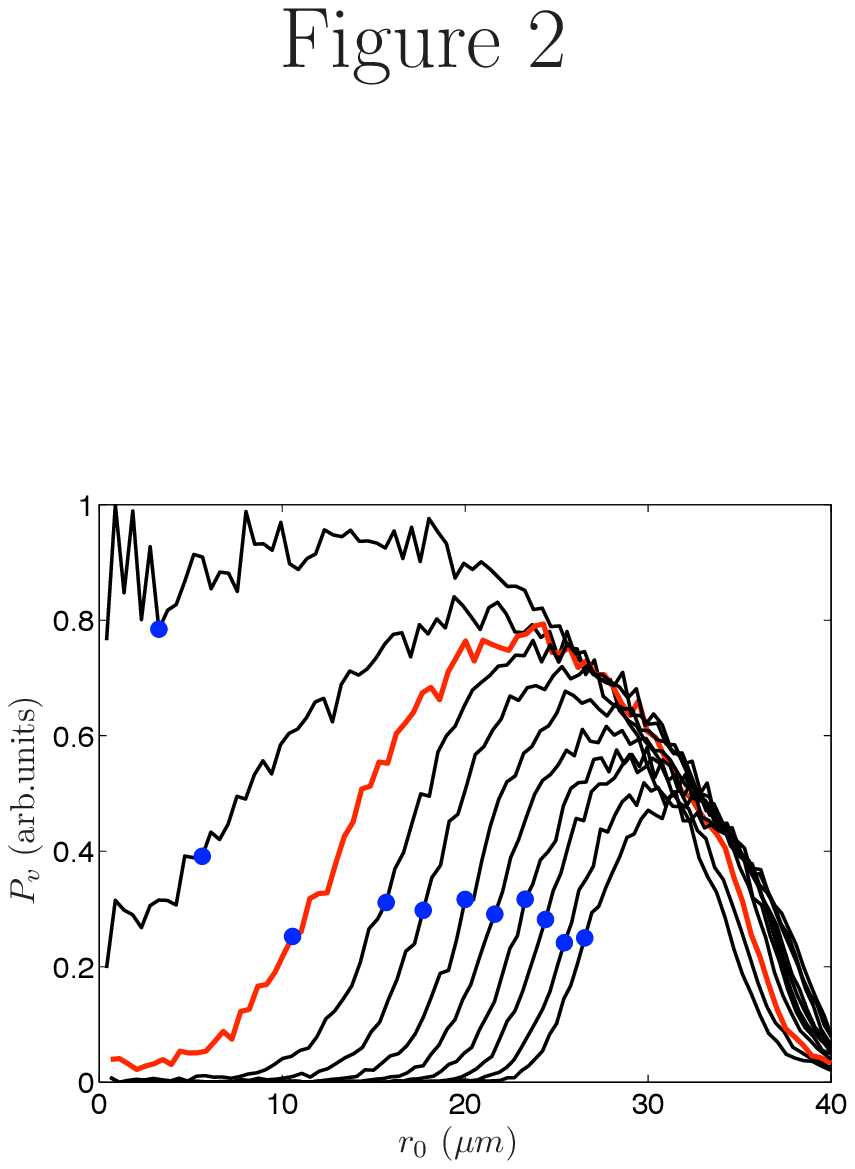}
\caption{(Color online) Radial vortex occupation probability density at different temperatures (from left to right) $\{T_i\}= \{168, 158, 155, 148, 143, 138, 135, 131, 126, 120, 115\}$ nK.  The red line indicates the superfluid crossover at $T_{\rm sf}=155$ nK.}
\label{f2}
\end{figure}

To further quantify the emergence of vortices and antivortices due to the phase fluctuations, we have measured at each temperature point the probablility, $P_v(r_o)$, of finding a vortex or antivortex at radius $r_0$. This is done by locating all phase singularities in an instantaneous classical field configuration and averaging over 1000 different microstates. The classical field area is then divided into ellipsoidal strips of equal width and the vortex occupation probability is obtained by counting the number of phase singularities detected within each strip divided by the number of microstates sampled (this is the \emph{radial} averaging discussed earlier). Thus obtained probabilities for a set of temperatures, $\{T_i\}= \{168, 158, 155, 148, 143, 138, 135, 131, 126, 120, 115\}$ nK, are plotted as functions of radial distance, $r_0$, in Fig.~\ref{f2}. The bullets denote the coherence length, measured as $1/e$ radius of $g(0,r_0)$ (defined in Sec. C below) and the red curve is for $T_{\rm sf}=155 n$K.  There is a sudden jump in the vortex occupation probability at the cross-over. In the superfluid phase there is a vortex-free region at small radii. Vortex pairs are observed in a narrow band near the edge of the coherent region of the system, i.e. the spatial region that has a flat phase in Fig~\ref{f1}(c). Thus the system can be divided in three concentric regions: central coherent and vortex free BEC-like region, coherent BKT-like region where vortices are bound, and an incoherent thermal outmost region where vortices are free.     

\subsection{Condensate fraction}
\begin{figure}
\center
\includegraphics[viewport= 180 410 430 610,clip, width=8.6cm]{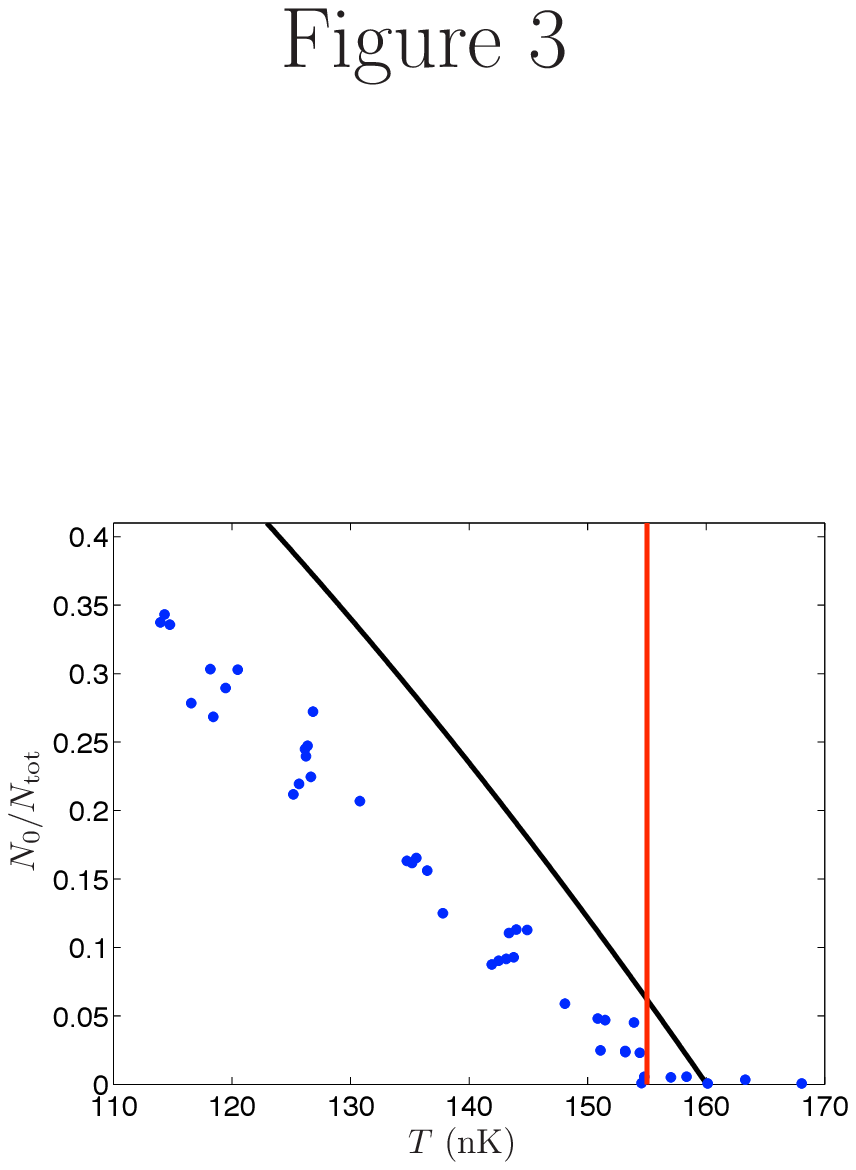}
\caption{(Color online) Condensate fraction as determined by diagonalization of the one-body density matrix as function of temperature. The markers are the data, the black curve is the 2D ideal gas result and the red vertical line indicates the superfluid crossover temperature $T_{\rm sf}$. }
\label{f3}
\end{figure}

We compute the one-body density matrix, Eq.~(\ref{eq1}), for the classical field by assuming ergodicity, which allows us to replace the ensemble average by a time average. The number of condensed particles, $N_0$, is obtained by computing the largest eigenvalue of the density matrix. Figure \ref{f3} displays the condensate fraction as a function of temperature. The curve is plotted to provide comparison with the pure 2D ideal gas relation
\begin{equation}
\frac{N_0}{N}=1-\left( \frac{T}{T_0}\right)^2,
\label{eq15}
\end{equation} 
where $T_0$ is the critical temperature calculated for a quasi-2D ideal-gas in our trap geometry containing $N_{\rm tot}=1.2\times10^5$ particles. The vertical line is the temperature $T_{\rm sf}=155$ nK, below which this system is superfluid and the markers are the simulation data. Unlike in the recent experiment of Kr\"uger \emph{et. al} \cite{Kruger2007a} we find the interactions to cause only a minor shift in the critical temperature from the non-interacting boson result.
In terms of the definition based on the eigenvalues of the density matrix, the system may be claimed to show Bose-Einstein condensation at all temperatures below $T_{\rm sf}$. Nevertheless it turns out that the system is best described in terms of phase fluctuating quasicondensate apart from the very lowest temperatures.  

\subsection{Coherence}
\begin{figure}
\center
\includegraphics[viewport= 180 410 430 610,clip, width=8.6cm]{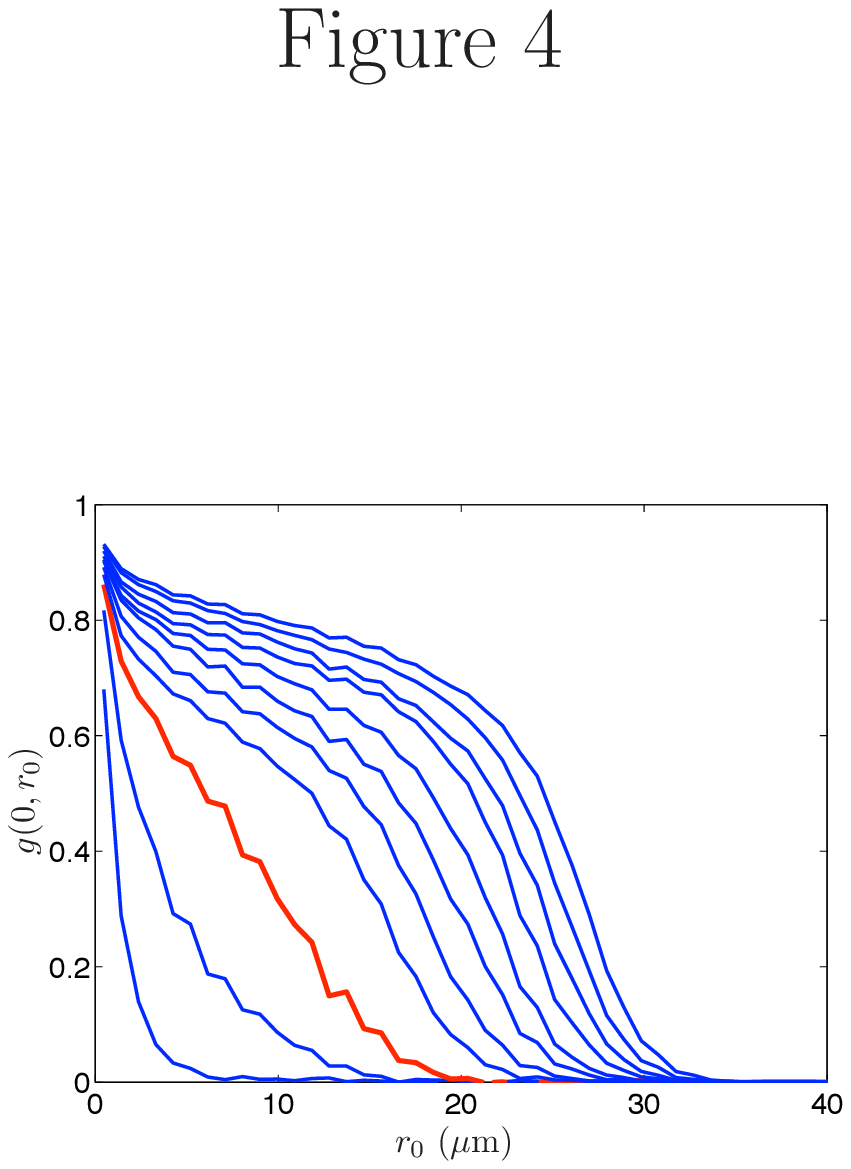}
\caption{(Color online) Two-point correlation functions, $g(0,r_0)$ as function of spatial distance, $r_0$, from the trap centre for a range of temperatures, $\{T_i\}$. The thicker red line is for the estimated superfluid cross-over temperature.}
\label{f4a}
\end{figure}

Coherence is an essential feature of superfluidity. The density matrix, Eq.~(\ref{eq1}), provides an useful probe for the global coherence between two spatially separated points in the system. In particular it conveys the knowledge of the correlation length and the information on the possible presence of long-range order. Figure~\ref{f4a} shows the two-point function, 
\begin{equation}
g(0,r_0)= \frac{\langle\Phi^*(0)\Phi(r_0)     \rangle}     { \sqrt{ \langle|\Phi(0)|^2 +n_{\rm th}(0) \rangle \langle |\Phi(r_0)|^2 +n_{\rm th}(r_0) \rangle} }
\label{eqg1}
\end{equation}
for different temperatures. For low temperatures, $T<T_{\rm sf}$, and small radii we witness power-law decay of $g(0,r_0)$ in accordance with Eq.~(\ref{eq5}), where as for temperatures, $T>T_{\rm sf}$, and/or near the edge of the coherent region, exponential decay is observed. The qualitative behavior changes at the cross-over temperature, $T_{\rm sf}$, denoted by the red line.

\begin{figure}
\center
\includegraphics[viewport= 180 410 430 610,clip, width=8.6cm]{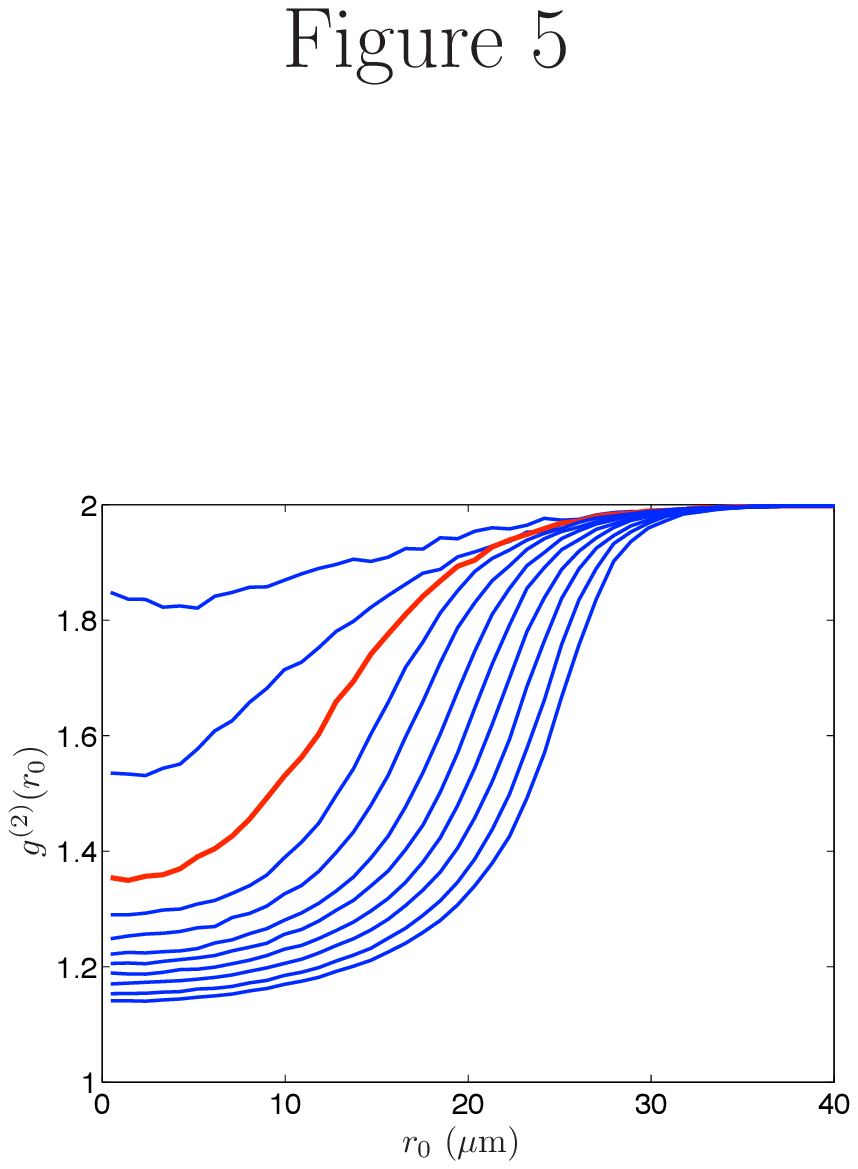}
\caption{(Color online) Second-order coherence functions $g^{(2)}(r_0)$ as function of spatial distance, $r_0$ from the trap centre for a range of temperatures, $\{T_i\}$. The thicker line is for the estimated superfluid cross-over temperature.}
\label{f4b}
\end{figure}

The second-order coherence function, Eq.~(\ref{eq6}), measures local coherence in the gas. Particularly, for a purely thermal sample $g^{(2)}_t(\bfr)=2$ and for completely coherent state $g^{(2)}_c(\bfr)=1$. In our inhomogeneous system  $g^{(2)}(\bfr)$ interpolates between these two values as shown in Fig.~\ref{f4b} where $g^{(2)}(r_0)$ is displayed for different temperatures as functions of the radial distance from the trap centre.  At the lowest temperatures, $g^{(2)}(r_0)$, shows a flat part in accordance with the presence of a nearly phase coherent BEC. At large enough radii and for high enough temperatures $g^{(2)}(r_0)$ approaches its thermal value. The plateau disappears near the temperature where the macroscopic wave function vanishes. The thicker red line indicates $g^{(2)}(r_0)$ for a result with $T\simeq T_{\rm sf}$. 

\begin{figure}
\center
\includegraphics[viewport= 180 410 430 610,clip, width=8.6cm]{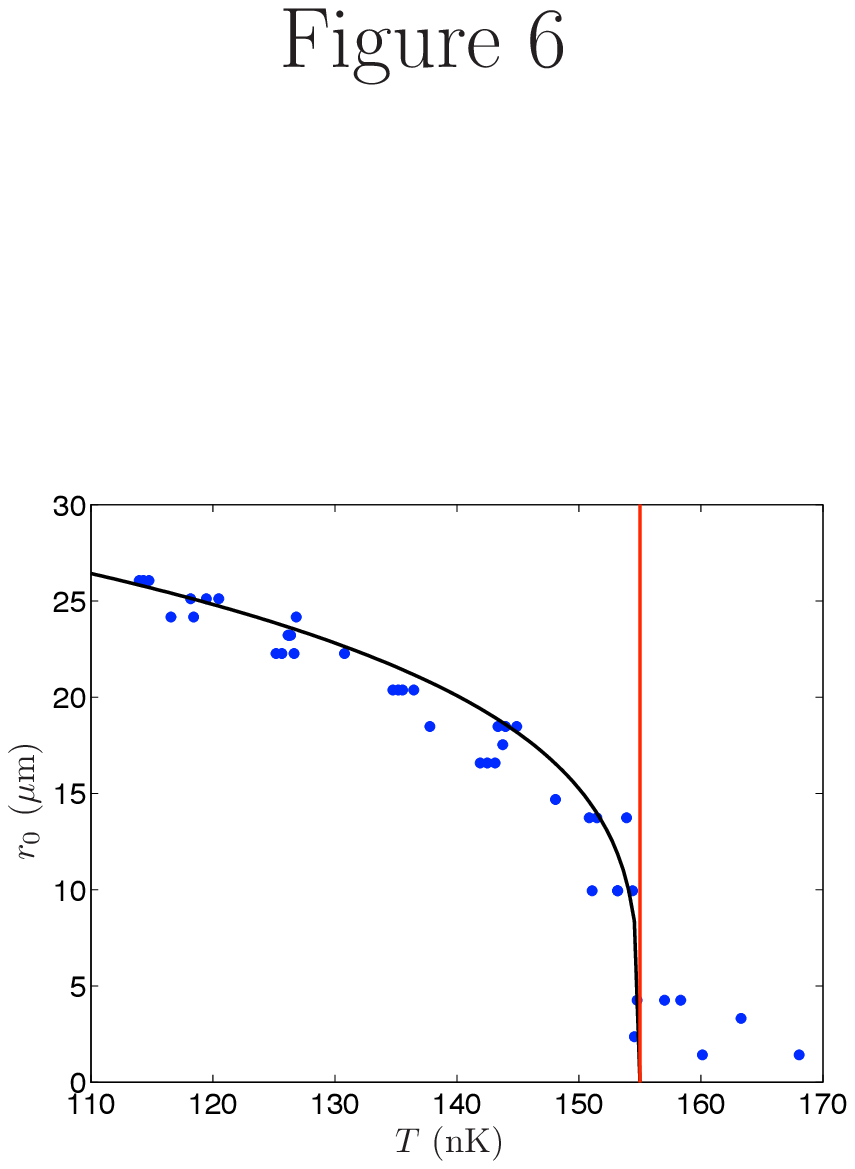}
\caption{(Color online) Correlation length as a function of temperature. The bullets are the numerical data, the vertical red line denotes the superfluid cross-over temperature and the black curve is the power-law function described in the text.}
\label{f5}
\end{figure}

In Fig.~\ref{f5}, we have plotted the coherence length as function of temperature. The values are obtained by measuring the $1/e$ width of the two-point correlation function. We have also plotted a function, $\xi_-(T) =|t|^p$, where, $t\propto (T_{\rm sf}-T)/T_{\rm sf}$, (solid curve) for $p=0.25$ and $T<T_{\rm sf}$. The value for the exponent, $p$, may be crudely explained in terms of the Thomas-Fermi radius, $R_{\rm TF} \propto N_0^{1/4}$ of an isotropic 2D condensate, since $N_0/N_{\rm tot}$ varies linearly with the temperature in the vicinity of the cross-over point, $T_{\rm sf}$. 

\subsection{Scissors modes}
\begin{figure}
\center
\includegraphics[viewport= 180 410 430 610,clip, width=8.6cm]{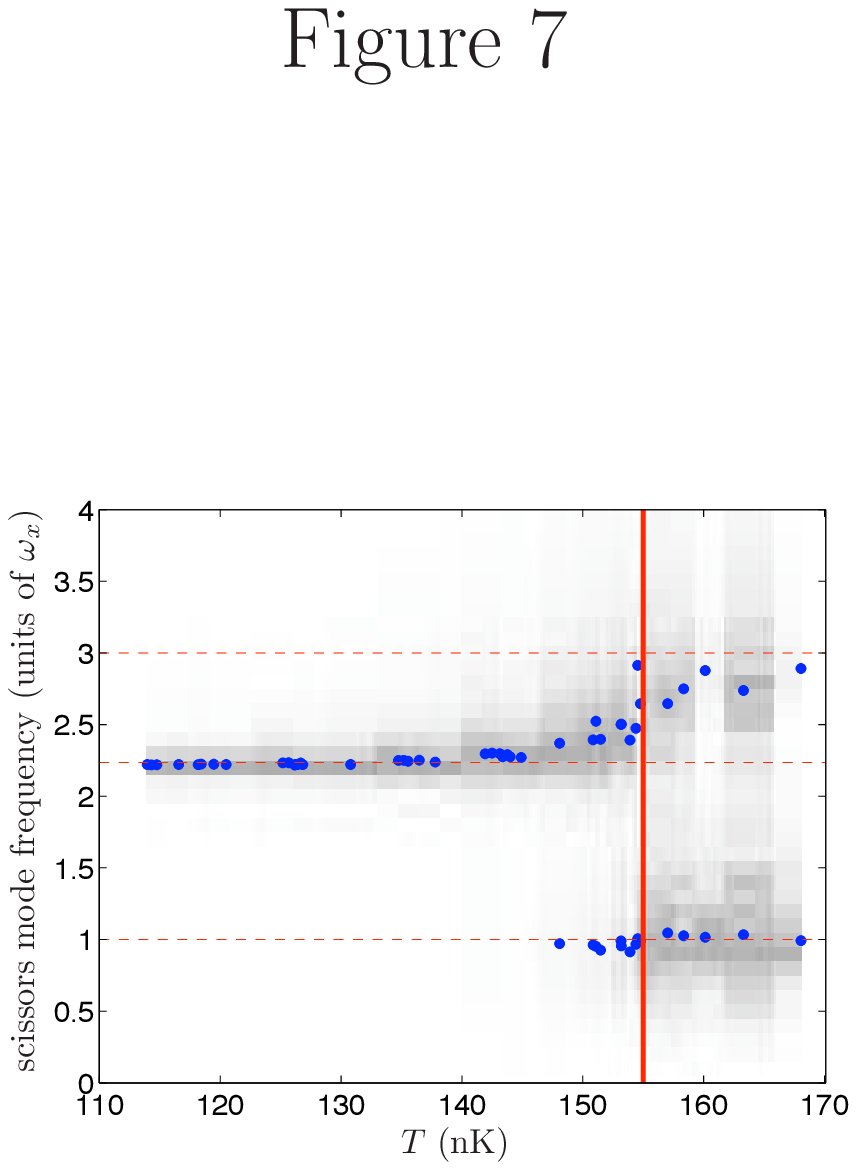}
\caption{(Color online) Scissors mode frequencies as a function of temperature. The horizontal dashed lines are the analytical predictions of Eqs.~(\ref{eq11}) and (\ref{eq12}) and the solid vertical line is our estimate of the superfluid transition temperature. The horizontal axis refers to the equilibrium temperature prior to the rotation of the state.  The normalised Fourier spectrum at each temperature is indicated in gray in the background.}
\label{f7}
\end{figure}

We have computed the scissors mode frequencies according to the description in the methods section. The obtained oscillation frequencies and their relative intensities are displayed in Fig.~\ref{f7}. The blue bullets are the mean oscillation frequencies obtained by fitting double Gaussian functions to the normalised Fourier spectrum at each temperature, which is indicated in gray in the background. The horizontal dashed red lines are the analytical predictions Eqs.~(\ref{eq11}) and (\ref{eq12}).  Above the cross-over temperature, $T_{\rm sf}$, indicated by the red vertical line, we obtain signal for two thermal modes whose frequencies are found to agree with the predictions of Eq.~(\ref{eq11}). At temperatures well below the cross-over, only one scissors mode persists with a frequency corresponding to that of a superfluid, see Eq.~(\ref{eq12}). This is the key feature verifying the quasicondensate to be superfluid. 

It is interesting to notice that we only observe two different scissors modes at all temperatures. The upper `irrotational' mode simply experiences frequency shift across the cross-over associated with the change in superfluid density of the system.  This is to be contrasted with  3D systems where in general three different scissors modes exist and the superfluid scissors mode experiences downward (as opposed to the upward shift seen in Fig.~\ref{f7}) frequency-shift on increasing temperature across the cross-over \cite{Marago2001a}.

We have also computed the scissors modes for different systems of varying particle interaction strengths which could be experimentally realized using Feshbach resonances. We have verified that in the strongly interacting systems only a single scissors mode survives at all temperatures, making it difficult to distinguish the superfluid and thermal response of the system from one another.

\section{Discussion}

In conclusion, we have studied the problem of superfluid-normal cross-over in a real, trapped quasi-2D Bose gas. In such systems the formation of pure Bose-Einstein condensation is challenged by the long-wavelength phase fluctuations and this fact has made the characterization of such systems difficult both theoretically and experimentally. By performing classical field simulations for these systems, we have shown that such quasicondensates are superfluid below the cross-over temperature $T_{\rm sf}$. This conclusion is based on observations of the global coherence properties and scissors mode excitations of the system, which constitute the two major results of this article proving superfluidity of quasicondensates. In particular, the emergence of condensate scissors-mode below the cross-over temperature provides an unequivocal microscopic evidence of non-classical rotational inertia and thus the superfluidity of the system. We have not found signs of fragmentation in terms of the eigenvalues of the density matrix below the cross-over temperature. A similar conclusion is obtained from the fact that only a single condensate peak is observed in  momentum space. This indicates that despite of the prevailing phase fluctuations, the superfluid state of these systems resembles more closely that of a single-mode trapped Bose-Einstein condensate than the bulk Berezinskii-Kosterlitz-Thoules superfluid phase.

Quasi-2D quantum gases are currently under active experimental investigation. While experiments have verified aspects of first-order coherence in this system consistent with the Berezinskii-Kosterlitz-Thouless transition, there has yet to be any direct evidence of superfluidity. In this paper we have reported a series of tests that can be realized in an experiment and should cast light on the formation of superfluidity in these systems.

\begin{acknowledgments}
This research was supported by the Marsden Fund of New Zealand, the University of Otago, and the New Zealand Foundation for Research Science and Technology under the contract NERF-UOOX0703: Quantum Technologies. MJD acknowledges the financial support of the Australian Research Council and the Queensland State Government.
\end{acknowledgments}

\end{document}